**Schedule-based Analysis of Transmission Risk in Public Transportation Systems**

**Jiali Zhou, Corresponding Author**

Department of Civil and Environmental Engineering

Northeastern University, Boston MA 02115

Tel: 412-708-2493; Email: zhou.jiali1@northeastern.edu

**Haris N. Koutsopoulos**

Department of Civil and Environmental Engineering

Northeastern University, Boston MA 02115

Tel: 617-373-6263; Email: h.koutsopoulos@northeastern.edu

Declarations of interest: none

**ABSTRACT**

Airborne diseases, including COVID-19, raise the question of transmission risk in public transportation systems. However, quantitative analysis of the effectiveness of transmission risk mitigation methods in public transportation is lacking. The paper develops a transmission risk modeling framework based on the Wells-Riley model using as inputs transit operating characteristics, schedule, Origin-Destination (OD) demand, and virus characteristics. The model is sensitive to various factors that operators can control, as well as external factors that may be subject of broader policy decisions (e.g. mask wearing). The model is utilized to assess transmission risk as a function of OD flows, planned operations, and factors such as mask-wearing, ventilation, and infection rates. Using actual data from the Massachusetts Bay Transportation Authority (MBTA) Red Line, the paper explores the transmission risk under different infection rate scenarios, both in magnitude and spatial characteristics. The paper assesses the combined impact from viral load related factors and passenger load factors. Increasing frequency can mitigate transmission risk, but cannot fully compensate for increases in infection rates. Imbalanced passenger distribution on different cars of a train is shown to increase the overall system-wide infection probability. Spatial infection rate patterns should also be taken into account during policymaking as it is shown to impact transmission risk. For lines with branches, demand distribution among the branches is important and headway allocation adjustment among branches to balance the load on trains to different branches can help reduce risk.

**Keywords:** COVID-19, Transmission risk, Public transportation, Mask-wearing, Ventilation, Operating strategies

1. Introduction

COVID-19, an infectious disease that continues to spread around the world despite development of effective vaccines, has far-reaching societal impacts (Coronaviridae Study Group of the International Committee on Taxonomy of Viruses, 2020). Public officials have taken various measures to control the spread of the disease, such as closing non-essential businesses and facilities, encouraging non-essential workers to stay home, and mandating mask wearing (Humphreys, 2020). Public transportation and its role in COVID-19 spreading has been part of the discussion.

Various studies in the literature have examined the impact of pandemics on public transportation ridership, attitudes towards using public transportation, etc. As a result, and in combination with local directives and business shut-down mandates, a dramatic decrease in transit ridership occurred in all metropolitan areas at the early stages of COVID-19 (Dickens, 2020; Jenelius and Cebecauer, 2020; Liu et al., 2020; Orro et al., 2020; Park, 2020).

Researchers have also studied the role of public transportation systems in the transmission of airborne diseases, and found correlations between influenza-like or COVID-19 illnesses with public transportation systems (Goscé and Johansson, 2018; Harris, 2020). Researchers also analyzed various cases of COVID-19 transmission in bus trips to directly assess the risk of transmission in public transportation (Luo et al., 2020; Shen et al., 2020). Zhu et al. (2021) studied various travel modes and travel destinations and the impacts on COVID-19 transmission in different cities and found that different travel modes have impacts on the number of cases in the cities.

Several studies applied and adapted the concept of contact networks and the "susceptible-infectious-recovered" (SIR) framework for transmission modeling of COVID-19 and other airborne diseases (Chen et al., 2020; Prem et al., 2020). Mo et al. (2021) developed a public transit contact network and found that partial closure of bus routes cannot fully contain the spreading of an epidemic. These studies focus on the dynamic evolution of transmission risk in the general population of a large metropolitan area, using a time horizon of days or months. They are dynamic, modeling the evolution of the spread of the disease over time but do not provide detailed accounting of how everyday operations of public transportation systems contribute to the transmission, and are not sensitive to the operational characteristics of transit services.

Other studies aim to assess the risks in small, confined spaces in relatively short time periods, normally within a few hours. For such cases, Riley et al. (1978) proposed a model and used it for the epidemiological study of a measles outbreak. The model, known as the Wells-Riley model, has been used extensively by researchers to study the transmission risk of various respiratory diseases, including influenza, SARS, etc. inside hospital wards, classrooms, and transit vehicles (Andrews et al., 2013; Barnett, 2020; Chen et al., 2011; Dai and Zhao, 2020; Fennelly and Nardell, 1998; Furuya, 2007; Ko et al., 2004; Rudnick and Milton, 2003; Stephens, 2012). The original model estimates the expected number of infections as a function of the ventilation rate, exposure time, number of carriers, and the infectiousness of the disease. The studies that looked at transmission risk in transit vehicles treated public transportation vehicles as a close-to-static indoor environment, assuming a fixed number of people and infectors (Andrews et al., 2013; Barnett, 2020; Chen et al., 2011; Dai and Zhao, 2020; Fennelly and Nardell, 1998; Furuya, 2007; Ko et al., 2004; Rudnick and Milton, 2003; Stephens, 2012). The exposure time is also assumed to be constant. As a result, they are more appropriate for intercity travel and do not capture the temporal and spatial characteristics of public transportation demand in urban areas.

Recently, Zhou and Koutsopoulos (2021) proposed a modified Wells-Riley model for risk analysis in public transportation systems that captures passenger flow spatio-temporal characteristics and actual operations. The model is utilized to assess overall risk as a function of passenger flows, actual operations, and factors such as mask-wearing, ventilation, infectiousness of disease, and infection rate in the population. The model is integrated with a microscopic simulation model of subway operations (SimMETRO) (Koutsopoulos and Wang, 2007; Zhou et al., 2020) for train trip-level analysis of transmission risk. While the model is sensitive to the various factors that impact risk, its reliance on a detailed simulation model for trip-based analysis limits its applicability.

The objective of this study is to present a simplified model, analogous to a planning tool, that can quickly analyze and quantify transmission risk under various operating strategies and other risk mitigation approaches so that decision makers can make informed decisions.

The main contribution of this study is the development and application of a quantitative framework for risk assessment that is sensitive to demand and service delivery in public transportation systems. It proposes a method to model the risk of transmission of airborne diseases based on the Wells-Riley model, and evaluates the transmission risks under planned service conditions, as opposed to trip-based conditions, using as inputs Automatic Fare Collection (AFC) and Automatic Vehicle Location (AVL) data. The framework can be utilized to evaluate the effectiveness of agency-controlled measures such as operating strategies and ventilation, as well as exogenous factors such as mask-wearing, infection rates in the population, and infectiousness of the disease. It is sensitive to the spatial characteristics of demand and infection rates, and applicable to any transit system.

The paper is organized as follows. Section 2 discusses the derivation of a Wells-Riley based model incorporating the demand and supply dynamics in a public transportation system in a general analytical framework. Section 3 introduces a case study using data from the Massachusetts Bay Transportation Authority (MBTA) Red Line. Section 4 discusses the results from the case study and evaluates the effectiveness of different mitigation strategies. Section 5 concludes the paper.

## 2. Methodology

### 2.1 The Wells-Riley Model

Riley et al. (1978) proposed the Wells-Riley model for infection probability evaluation in indoor environments, considering the intake dose of airborne pathogens in terms of the number of infectious particles.

$$P = 1 - \exp\left(-\frac{Ipqt}{Q}\right) \tag{1}$$

where $P$ is the probability of infection, $I$ the number of infectors (carriers), $p$ the breathing rate per person (m$^3$/hour), $q$ the quanta generation rate (quanta/hour), $t$ the exposure time (hour), $Q$ the room ventilation rate of clean air (m$^3$/hour).

It is worth mentioning that $q$, the quanta generation rate (quanta/hour), is not a physical unit. Instead, it captures the average number of infectious particles that are exhaled to the environment and the infectiousness of the particles. The value of $q$ can be back-calculated from epidemiological studies of outbreak cases using equation (1) (Rudnick and Milton, 2003; Stephens, 2012).

To incorporate additional factors, including mask-wearing and mask filtration efficiency, the original model (equation (1)) has been modified and further developed in the literature (Fennelly and Nardell, 1998; Furuya, 2007; Zhou and Koutsopoulos, 2020). Considering the environment of a train cabin, with both a fraction of infectors and susceptible persons wearing a mask, the probability of infection can be expressed as:

$$P = 1 - \exp\left(-\frac{I(F_m \times p)(R_m \times q)t}{Q}\right) \tag{2}$$

The infectious particles exhaled by infectors are filtered at a certain rate, $R_m$, the particle penetration rate of the mask. $R_m = 1$ represents the case of no mask wearing where all infectious particles from the infector are exhaled into the air; $R_m = 0$ means a mask with 100% particle blockage. Similarly, the particles inhaled by susceptible persons could be filtered at a certain rate $F_m$ ($F_m = 1$ means no mask, and $F_m = 0$ 100% particle blockage).

### 2.2 Modeling Infection Risk for Transit Operations

Let us consider a passenger who travels from station $i$ to station $j$. During their trip they interact with passengers traveling from $r$ to $s$ if $r<j$ and $s>i$. Conditional on the number of infectors $n_{rs}$ from station $r$ to station $s$, the probability that a passenger from $i$ to $j$ is not infected by the $n_{rs}$ infected passengers who travel from $r$ to $s$ is given by:

$$P_{ij}^{rs}(not\ infected\,|\,n_{rs}) = \exp\left(-\frac{n_{rs}qpt_{ij}^{rs}}{Q}\right) \tag{3}$$

where $t_{ij}^{rs}$ is the time a passenger from $i$ to $j$ is exposed to passengers from $r$ to $s$ (based on the overlap between the two OD pairs).

The number of infectors on OD pair $rs$, assuming that the probability of a person being infected is $\pi$ (infection rate), follows the binomial distribution with parameters $N_{rs}$ (the total number of passengers traveling from $r$ to $s$) and $\pi$. Thus, the probability that a passenger traveling on OD pair $ij$ will not be infected by passengers from station $r$ to $s$ is given by:

$$P_{ij}^{rs}(not\ infected) = \sum_{n_{rs}=0}^{N_{rs}} \exp\left(-\frac{n_{rs}qpt_{ij}^{rs}}{Q}\right)\binom{N_{rs}}{n_{rs}}\pi^{n_{rs}}(1-\pi)^{N_{rs}-n_{rs}}] \tag{4}$$

$n_{rs}$ is the number of infectors (carriers). $N_{rs}$ is the number of passengers aboard a train from station $r$ to station $s$, $\pi$ the infection rate in the population. $\pi$ can have different values for passenger groups depending on their origin/destination, in order to accurately capture the spatial pattern of the spread of the virus.

For large values of $N_{rs}$ and small $\pi$, the binomial distribution can be approximated using the Poisson distribution:

$$P(n_{rs}) = \binom{N_{rs}}{n_{rs}}\pi^{n_{rs}}(1-\pi)^{N_{rs}-n_{rs}} \approx \exp(-N_{rs}\pi)\frac{(N_{rs}\pi)^{n_{rs}}}{n_{rs}!} \tag{5}$$

Hence, the probability that a passenger traveling from $i$ to $j$ will not be infected by passengers traveling from $r$ to $s$ is given by:

$$P_{ij}^{rs}(not\ infected) = \sum_{n_{rs}=0}^{N_{rs}} \exp\left(-\frac{n_{rs}qpt_{ij}^{rs}}{Q}\right)\exp(-N_{rs}\pi)\frac{(N_{rs}\pi)^{n_{rs}}}{n_{rs}!}] \tag{6}$$

Assuming independence from station to station, the probability $P_{ij}$ of a passenger traveling on the OD pair $ij$ being infected is:

$$P_{ij} = 1 - \prod_{rs}[\sum_{n_{rs}=0}^{N_{rs}}[\exp\left(-\frac{n_{rs}qpt_{ij}^{rs}}{Q}\right)\exp(-N_{rs}\pi)\frac{(N_{rs}\pi)^{n_{rs}}}{n_{rs}!}] \tag{7}$$

Following Zhou and Koutsopoulos (2021), the model can be extended to accommodate mask-wearing behavior. Let $f_m$ be the fraction of mask-wearing passengers. The probability $\overline{P_{ij}^m}$ of a susceptible person traveling from $i$ to $j$ and wearing a mask not infected by (masked and non-masked) infectors is calculated by combining equations (5), (6) and (7):

$$\overline{P_{ij}^m} = \prod_{rs}[\sum_{n_{rs}=0}^{N_{rs}} \exp\left(-\frac{n_{rs}f_m(F_m \times p)(R_m \times q)t_{ij}^{rs}}{Q}\right)\exp(-N_{rs}\pi)\frac{(N_{rs}\pi)^{n_{rs}}}{n_{rs}!}] \times$$

$$\prod_{rs}\left[\sum_{n_{rs}=0}^{N_{rs}} \exp\left(-\frac{n_{rs}(1-f_m)(F_m \times p)(q)t_{ij}^{rs}}{Q}\right)\exp(-N_{rs}\pi)\frac{(N_{rs}\pi)^{n_{rs}}}{n_{rs}!}\right] =$$

$$\prod_{rs}\left[\sum_{n_{rs}=0}^{N_{rs}} \exp\left(-\frac{n_{rs}f_m(F_m \times p)(R_m \times q)t_{ij}^{rs}+n_{rs}(1-f_m)(F_m \times p)(q)t_{ij}^{rs}}{Q}\right)\exp(-N_{rs}\pi)\frac{(N_{rs}\pi)^{n_{rs}}}{n_{rs}!}\right] \tag{8}$$

$n_{rs}f_m$ is the masked proportion of $n_{rs}$. $R_m$, as stated earlier, is the effectiveness of a mask.

Similarly,

$$\overline{P_{ij}^{n\_m}} = \prod_{rs}\left[\sum_{n_{rs}=0}^{N_{rs}} \exp\left(-\frac{n_{rs}f_m(1 \times p)(R_m \times q)t_{ij}^{rs} + n_{rs}(1-f_m)(1 \times p)(q)t_{ij}^{rs}}{Q}\right)\exp(-N_{rs}\pi)\frac{(N_{rs}\pi)^{n_{rs}}}{n_{rs}!}\right] \quad (9)$$

where $\overline{P_{ij}^{n\_m}}$ is the probability that a non-masked susceptible passenger traveling from station $i$ to station $j$ is not infected.

Consequently, the probability of infection for a passenger traveling from $i$ to $j$ is given by:

$$P_{ij}^m = 1 - \overline{P_{ij}^m} \quad (10)$$

$$P_{ij}^{n\_m} = 1 - \overline{P_{ij}^{n\_m}} \quad (11)$$

Using equations (8), (9), (10), and (11), $r_{ij}$, the expected number of infected passengers traveling on OD pair $ij$ per trip is given by:

$$r_{ij} = P_{ij}^{n\_m}(1-f_m)\,D_{ij} + P_{ij}^m f_m\,D_{ij} \quad (12)$$

$D_{ij}$ is the OD flow *from i to j* (susceptible).

Based on the above equations, the expected number of infections at the system level, *r*, and the overall probability of infection can be calculated taking into account (weighted by) the distribution of the trips in the system. These metrics are useful measures of risk and can be used to assess the impact of operating characteristics, as well as the effectiveness of different mitigation strategies related to mask wearing, ventilation, etc., on a consistent basis.

Figure 1 summarizes the risk calculation approach as a function of service characteristics and virus related parameters. The model receives as input the timetable, OD demand matrix, and train travel times between stations. Given the fraction of infectors in the general population (could vary by station) and the fraction of mask-wearing passengers, the model calculates the infection risk.

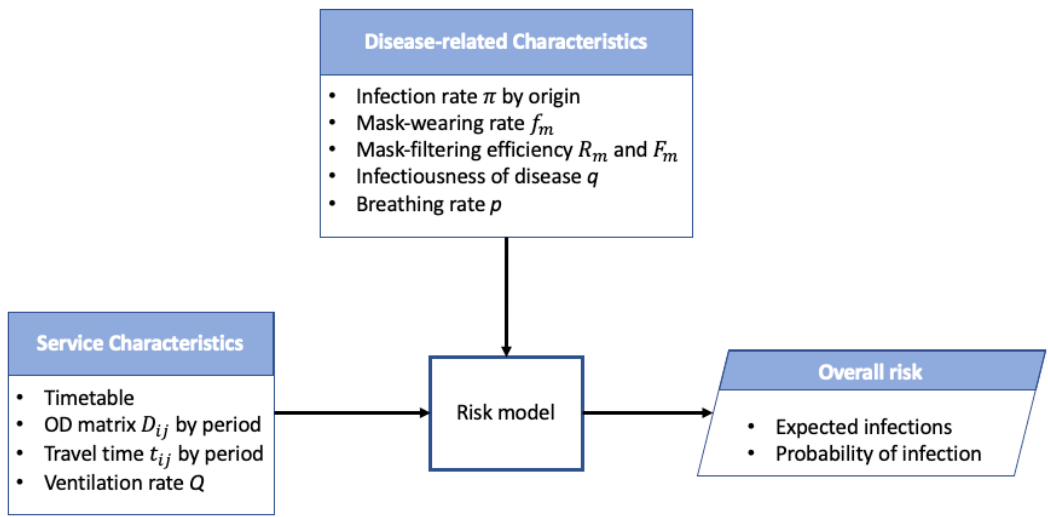

Figure 1 Transmission risk assessment approach

The impact from factors such as mask-wearing, ventilation rates, travel times, OD demand levels, especially the trade-offs between them, can provide interesting insights for policymakers with regard to what can be implemented to lower transmission risk in public transportation systems. Similarly, the importance of virus related characteristics and their interactions with service characteristics are of interest. These factors can be categorized into two broad categories: viral load and passenger load factors. Factors directly impacting the viral load are mask-wearing, mask filtration efficiency, ventilation, and OD travel times. Demand (OD flows) and frequency of service are passenger load related factors.

In the discussion that follows, the superscript $^o$ is used to indicate base conditions in terms of service frequency, demand (OD flows), ventilation, infectiousness, travel time, and mask filtration efficiency. Let $q^o$ be the base quanta generation rate (infectiousness of disease), $h^o$ the base case headway, $D^o$ the base demand rate (passengers per unit time) in terms of OD flows, $N_{rs}^o$ the base case average number of passengers traveling from $r$ to $s$ per trip, $Q^o$ the base case ventilation rate, and $t_{ij}^{rs o}$ the base case travel time a passenger from $i$ to $j$ is exposed to passengers traveling from $r$ to $s$. Factors $\alpha$, $\beta$, $\gamma$, $\delta$, and $\varepsilon$ are used to adjust the base values.

$$q = \alpha q^o \tag{13}$$

$$h = \beta h^o \tag{14}$$

$$D = \gamma D^o \tag{15}$$

$$t_{ij}^{rs} = \delta t_{ij}^{rs o} \tag{16}$$

$$Q = \varepsilon Q^o \tag{17}$$

Consequently, $N_{rs}$, the average number of passengers per trip from $r$ to $s$, as a function of the corresponding average base number of passengers $N_{rs}^o$ per trip, is given by:

$$N_{rs} = \beta \gamma N_{rs}^o \tag{18}$$

The probability of infection (equation 10) can then be expressed as a function of the base case conditions and the scaling factors.

$$P_{i,j}^m$$

$$= 1 - \prod_{rs}\left[\sum_{n_{rs}=0}^{\beta\gamma N_{rs}^o} \exp\left(-\frac{n_{rs}(F_{m\times}p)\big(1-f_m(1-R_m)\big)(\alpha \times q^o)\delta t_{ij}^{rs o}}{\varepsilon Q^o}\right)\exp(-\beta\gamma N_{rs}^o \pi)\frac{(\beta\gamma N_{rs}^o \pi)^{n_{rs}}}{n_{rs}!}\right] =$$

$$1 - \prod_{rs}\left[\sum_{n_{rs}=0}^{BN_{rs}^o}\exp\left(-A\frac{n_{rs}(p)(q^o)t_{ij}^{rs o}}{Q^o} - BN_{rs}^o \pi\right)\frac{(BN_{rs}^o \pi)^{n_{rs}}}{n_{rs}!}\right] \tag{19}$$

where,

$$A = \frac{\alpha\delta}{\varepsilon}(1 - f_m(1-R_m))F_m \tag{20}$$

$$B = \beta\gamma \tag{21}$$

Meta-parameter $A$ represents the combined effect of the factors that affect the airborne viral load, and $B$ the factors that affect the passenger (carrier) load, as shown in Figure 2. They can be used to calculate the change in passenger load required to compensate for an increase in viral load, for example, due to an increase in the infectiousness of the disease. It should also be noted that $A$ is impacted mainly by external factors (e.g. mask-wearing), while $B$ is a function of demand characteristics and operating policy (e.g. frequency of service).

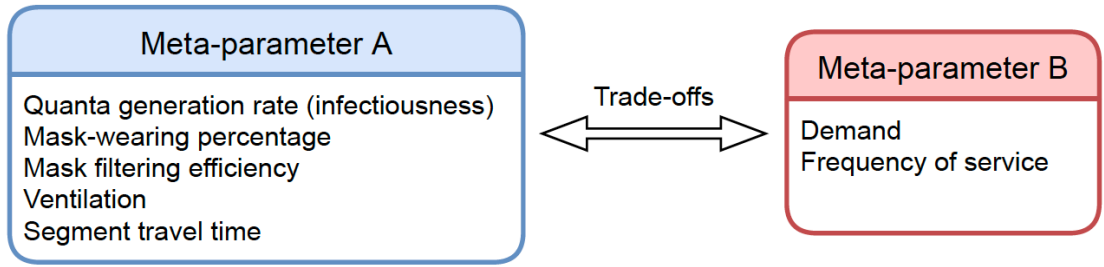

Figure 2 Meta-parameters $A$ and $B$

Trade-offs and interactions among factors related to meta-parameters $A$ and $B$, can also be explored. Agencies may be interested in maintaining a target risk level. As factors change, for example infectiousness of a virus due to new variants, equations (20) and (21) can be used to assess quickly the changes in operating strategies that may help mitigate the increase in risk. For example, $\alpha = 1.5$ (infectiousness increased by 50%) implies that A = 1.5 $A^0$ (the base value). Required changes in $F_m$ to reduce risk to the original level can then be assessed. Similarly, changes in headways (value of $\beta$) can also be assessed.

## 3. Application

The MBTA Red Line (Figure 3) provides the background for the case study. The Red Line is the highest frequency and ridership heavy rail transit line in the MBTA network with an average weekday ridership of 251,000 passengers (Massachusetts Bay Transportation Authority, 2019). The line is composed of a trunk section and two branches on the south end. The trunk section connects the north terminus (Alewife Station) to JFK/UMass station and is 14.2 km (8.8 miles) long with 13 stations. The Braintree branch is 14.2 km (8.8 miles) long with 5 stations and the Ashmont branch 4.7 km (2.9 miles) long with 4 stations.

The demand on the Red Line was greatly impacted by the COVID-19 pandemic as Figure 3 indicates. The analysis focuses on the evening peak period (16:00 to 18:30) using September 2020 demand levels (OD flows in 15-minute intervals) and the typical schedule. The direction we consider is the Southbound direction.

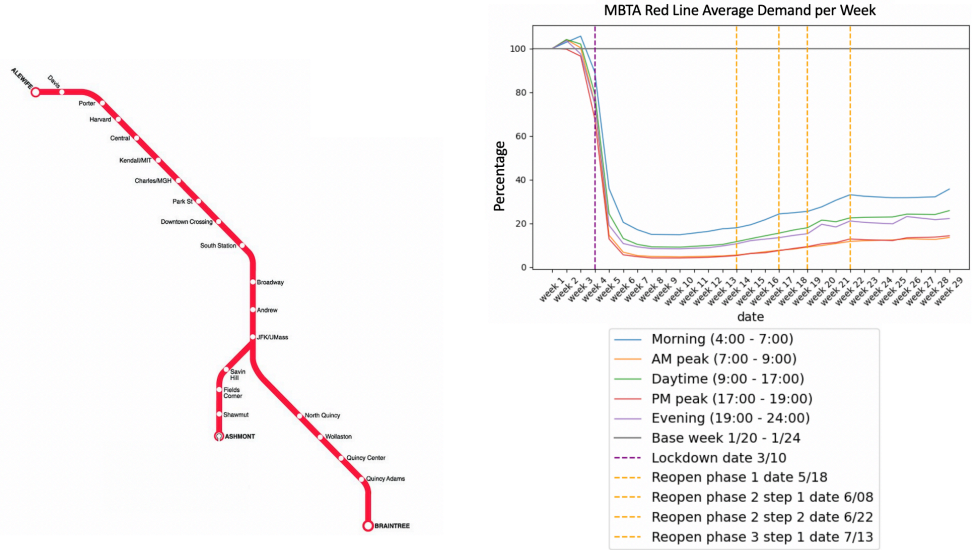

Figure 3 MBTA Red Line average demand per week after 2020/02/16

3.1 Experimental Design and Assumptions

Table 1 summarizes the base parameter values for the experiments.

Table 1 Base case parameters and assumptions

| 1 | Mask effectiveness in terms of particle penetration rate for exhaling (inhaling) $R_m = 0.5$ ($F_m = 0.5$), as used and reported in (Pan et al., 2020; Stutt et al., 2020), to model normal quality face coverings. |
|---|---|
| 2 | Mask-wearing percentage $f_m$ in the population is assumed to be 0%. |
| 3 | A $q$ value = 270 quanta/hour is assumed, which is also consistent with the findings from studies, as over 100 quanta/hour (Buonanno et al., 2020a; Buonanno et al., 2020b; Miller et al., 2021). The value is back-calculated using the method mentioned in studies (Rudnick and Milton, 2003; Stephens, 2012) by applying equation (1). Two COVID-19 outbreak cases in public transportation vehicles reported in the literature (Luo et al., 2020) are used to derive the $q$ value. |
| 4 | Ventilation rate in a cabin of a subway train $Q$ = 1958 m³/hour (Massachusetts Bay Transportation Authority, 2013). |
| 5 | Breathing rate per person (m³/hour), $p$ = 0.72 m³/hour (Environmental Protection Agency, 2015). |
| 6 | According to the published schedules, the scheduled headways from the branches are 9 minutes, coordinated to 4.5 minute headways in the trunk. |

7    Average infection (carrier) rates in the population are assumed to be 0.92% for all stations according to data as of June 2021 in the Greater Boston area.

8    The base case demand is assumed to be equal to September 2020 demand,

Based on the above assumptions, the base case system-wide infection probability is 2.8/1000. The base case assumes that passengers do not wear a mask ($f_m = 0\%$). For the calculation, we also assume that the air exchange by door-opening at stations is neglected and risk due to waiting at stations is not considered.

The objective of the case study is to apply the proposed model to assess transmission risk in subway systems and, explore the trade-offs between viral load and passenger load meta-parameters. The impact on transmission risk of headways, spatial patterns of infection rates (as opposed to assuming uniform rates across the service area), and passenger distribution on train cars, is also of interest.

## 4. Results and Discussion

### 4.1 Meta-parameters A and B trade-offs

The latest variants of COVID-19 are more infectious than the original strain, for example, the Delta variant is around two to five times as infectious as previous variants (Centers for Disease Control and Prevention, 2021; Herrero, 2021). When the infectiousness of the disease increases, in principle, the impact can be mitigated by changing other factors that contribute to the value of $A$, or by lowering the value of $B$ (representing passenger load) through changes in the frequency of service.

Figure 4a shows a heatmap of the system-wide probability of infection for unmasked susceptible passengers, for various values of the viral load meta-parameter $A$ and passenger load meta-parameter $B$. The lines represent the iso-risk curves for different values of meta-parameters $A$ and $B$. As expected, the system level overall infection probability increases as the values of $A$ and $B$ increase. However, meta-parameters $A$ and $B$ have different effects on transmission risk. Figure 4a can be a useful tool to quickly assess desired targets to mitigate impacts on risk due to changes in external factors. Let us consider the base case (overall infection probability = 2.8/1000) where $A = 0.5$ and $B = 1.0$ (point 1 in Figure 4a). A 20% increase in the virus infectiousness (for example due to mutation) means that $A = 0.6$ (point 2 in Figure 4a). The change in the $B$ parameter to mitigate the risk and bring it to previous levels requires passenger load changes that result in a new $B$ value of around 0.85 (point 3). Since $B$ is a function of frequency of service and OD flow patterns, and assuming that demand remains the same, headways have to change by a factor $\beta = 0.85$ (reduced) so that the risk returns to its original value. However, depending on the base operations, this may not be easily accomplished, given line (signal) capacity constraints. Alternatively, operators and policy makers, may attempt to mitigate the increased risk, due to the increased infectiousness, by changing other parameters that contribute directly to A. $\alpha$ is the factor that adjusts the scale for the base quanta generation rate $q^o$ ($q = \alpha q^o$), i.e. the infectiousness of the disease (it contributes to the value of $A$). To fully compensate the increase in the infectiousness, other factors contributing to the value of $A$ could also be changed to keep its value at the same level (assuming $B$ remains unchanged). For example, agencies can encourage mask wearing to increase the proportion $f_m$ in the population by distributing masks with better protection, such as N95 or surgical masks (Tirupathi et al., 2020). Figure 4b shows a heatmap of the system-wide probability of infection for masked susceptible passengers, for various values of the viral load meta-parameter $A$ and passenger load meta-parameter $B$. The results indicate that the infection risk is significantly lower for masked passengers compared to unmasked passengers.

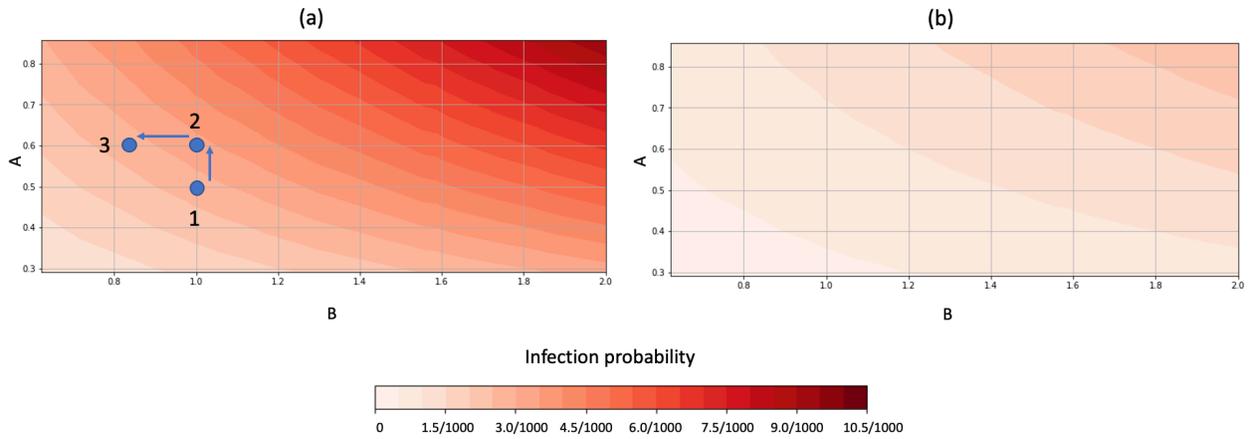

Figure 4 Infection probability iso-curves under various meta-parameter A and B values. (a) Infection probability for unmasked passengers; (b) Infection probability for masked passengers.

In the base case, $A = 0.5$. $A$ is a function of various viral load parameters ($A = \frac{\alpha\delta}{\varepsilon}(1 - f_m(1 - R_m))F_m$). Figure 5 illustrates the trade-offs among the factors contributing to meta-parameter $A$ so its value remains at $A = 0.5$. In particular, the figure illustrates the combination of values of $\alpha$ (infectiousness) and $F_m$ (mask effectiveness) and $\alpha$ and $f_m$ (mask wearing percentage) so that $A = 0.5$. Figure 5a shows the interactions between infectiousness ($\alpha$) and mask effectiveness ($F_m$). The results indicate that $F_m$ should decrease (mask efficiency should increase) as the value $\alpha$ increases. For example, a 2 times increase in infectiousness requires masks that can filter out approximately 75% of the particles (from the base case of 50%). Figure 5b shows the required increase in the proportion of mask-wearing passengers to compensate for an increase in infectiousness ($\alpha$) (so that the value of A remains the same at 0.5). A two times increase in infectiousness requires that mask wearing percentage increases to 100% (assuming mask-wearing effectiveness remains at 50%). However, increasing the mask-wearing proportion can compensate an at most 2 times increase in infectiousness. If the infectiousness increases by more than 2 times, mask wearing proportion alone cannot bring the overall risk back to the base level (even if all passengers wear a mask).

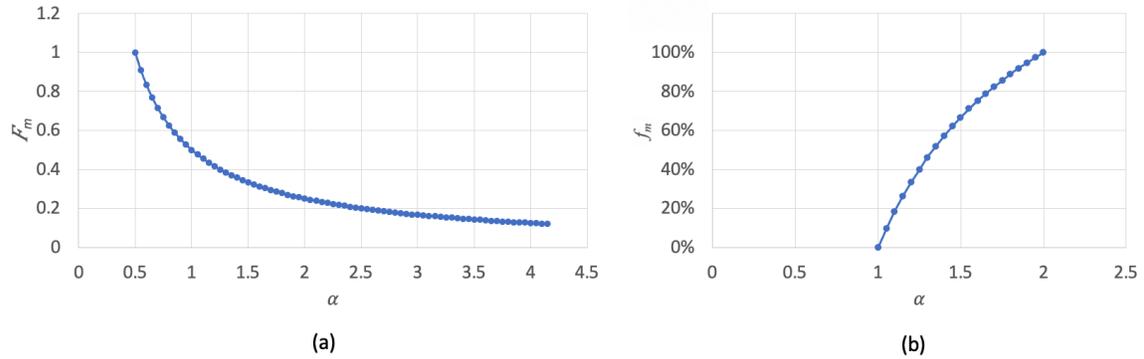

(a)                    (b)

Figure 5 Interactions between factors from meta-parameter $A$: (a) $\alpha$ vs. $F_m$, (b) $\alpha$ vs. $f_m$, (A = 0.5)

## 4.2 Impact of Operating Characteristics

It has been shown that operating strategies (headways) and service reliability impact the transmission risk (Zhou and Koutsopoulos, 2021). Headways impact the trainload (and passenger load related meta-parameter $B$) and subsequently the transmission risk. Figure 6 summarizes the impact on the system level risk of various infection (carrier) rates at different scheduled headways (x-axis). For other parameters not stated in the legends, the values are as stated in Table 1 (base case).

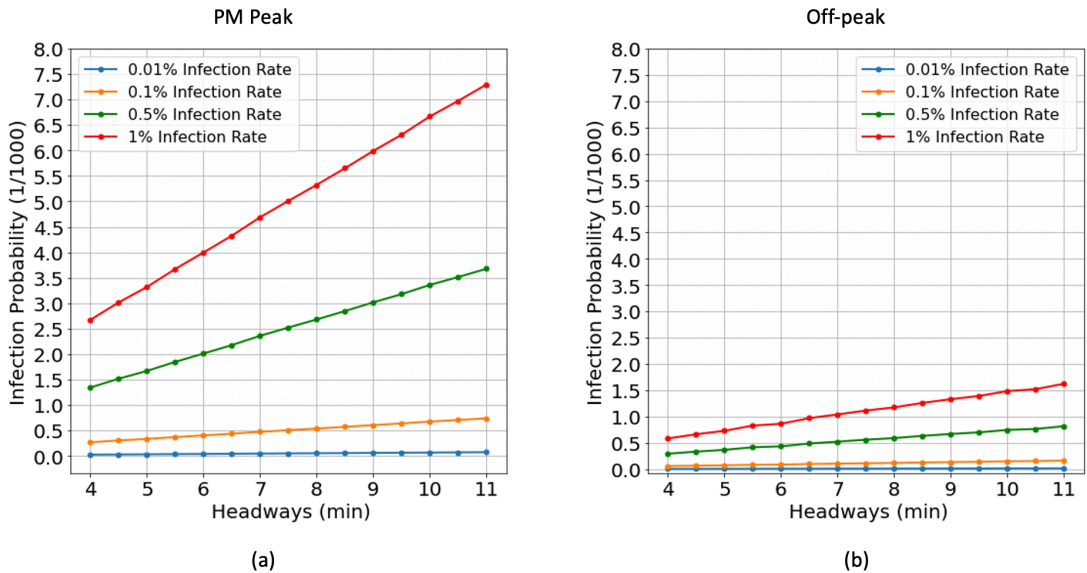

(a)                    (b)

Figure 6 Infection probability under different headways and infection (carrier) rates (a) PM peak. (b) Off-peak

Figure 6a shows the impact on the infection probability during the PM peak under various headways. When the population infection rate is lower than 0.1%, the risk is relatively low, and the differences among the various headway levels are small. As the infection rate increases from 0.1% to 1.0%, the risk increases significantly. Figure 6b shows the impact on the infection probability of various headways during off-peak hours. The average off-peak hourly demand is 32% of the peak hourly demand. Given the lower demand levels in the off-peak, the risk is much lower and relatively less sensitive to increases in infection rates compared to the PM peak case.

Figure 7 summarizes the impact on risk at the system level of mask-wearing at different headways (x-axis) during the PM peak. For other parameters not stated in the legends explicitly, the values are as stated in the base case.

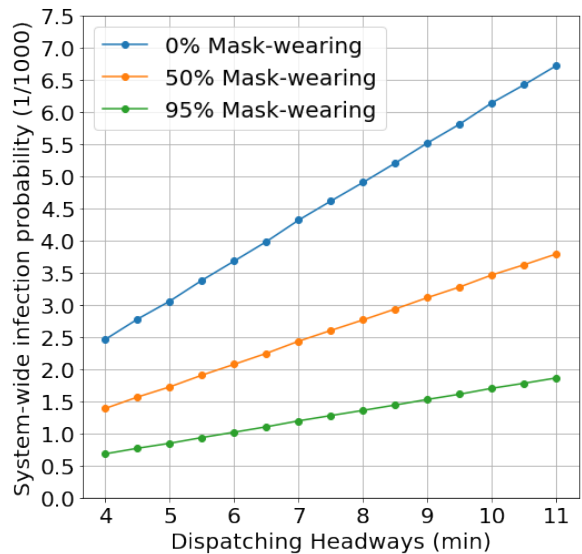

Figure 7 Infection probability under different headways and mask-wearing proportions during the PM peak

As shown in Figure 7, when passengers wear masks, the infection probability is not as sensitive to the service headways compared to the non-masked case. With 50% of passengers wearing masks, the infection probability decreases by approximately 44% from the base case. With 95% of passengers wearing masks, the infection probability decreases by approximately 73% from the base case. The mitigation effect of mask-wearing is significant.

Figure 8 shows a heatmap of the system-wide infection probability $P$ (expressed in units of 1 over 1000) under various combinations of headways and level of infectiousness. The results indicate that increasing the frequency of service can mitigate transmission risk and compensate for a rise in infectiousness of the disease. For example, when the headway is 9 min and the infectiousness is at the base case value, the infection probability is 5.52/1000. When the virus is 2 times more infectious than the base case, agencies need to adjust the headways to lower than 6.0 min to compensate for such a rise in infectiousness.

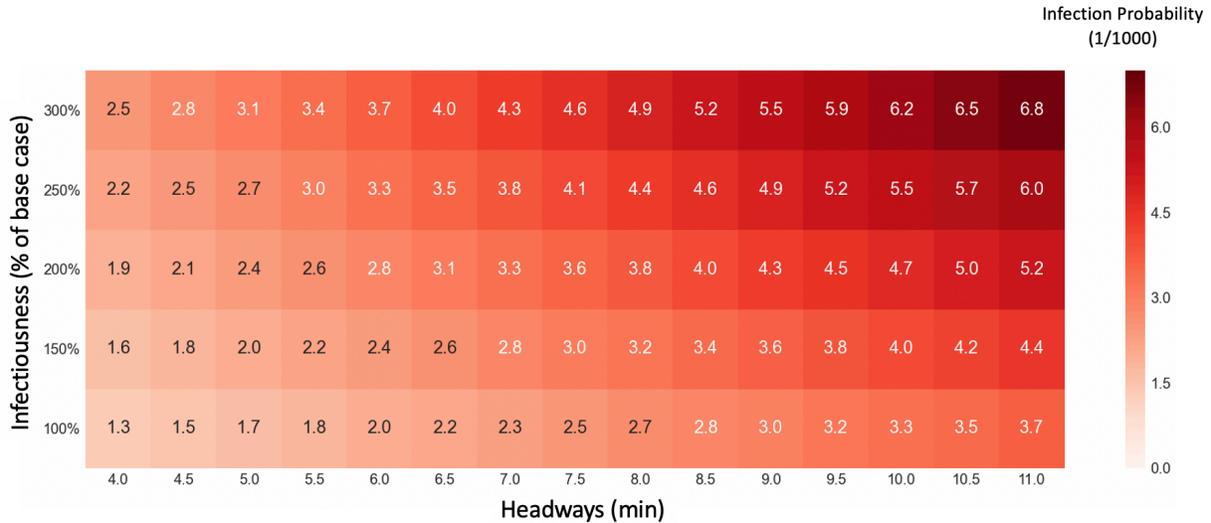

Figure 8 System-wide infection probabilities (in 1/1000) under various headways and infectiousness

When designing headways during a pandemic, it is also important to take the spatial demand patterns into account. In the MBTA Red Line case, the passenger demand travelling to different branches varies. Braintree (B) trains are carrying more passengers to their branch than Ashmont (A) trains. The Braintree branch is also longer in distance, so its passengers have longer exposure times. During normal operations, trains to Braintree and Ashmont dispatched at the trunk terminal (Alewife) alternate (A-B-A-B), and the headways between consecutive trains are equal. Since Ashmont demand is lower, equal headways between consecutive trains result in Ashmont trains having smaller loads compared to Braintree trains.

Using the transmission modeling framework developed in the previous section, we explore the impact of headway allocation between the branches on transmission risk. The settings are consistent with the base-case assumptions, except the headway allocation between branches. Let $H_{AB}$ be the headway of an Ashmont branch train with the Braintree train in front of it ($H_{BA}$ is the headway of a Braintree branch train with the Ashmont train in front of it). The headways in the branches are assumed to be the same as in the base case, 9 minutes. Under current practice, the headways in the trunk section are 4.5 minutes. Given the unequal demand of the two branches and in order to assess the impact of headway allocation, we assume that $H_{AB}$ ranges from 2.0 to 7.0 min, while $H_{BA}$ assumes a value of $9.0 - H_{AB}$ (which maintains the 9.0 minute headways in the branches).

Figure 9 illustrates the risk at the system and branch levels as a function of $H_{AB}$. The infection probability for Braintree trains decreases as $H_{AB}$ increases from 2.0 min to 7.0 min. At the same time, the infection probability for Ashmont trains increases. With the uneven allocation of headways, Braintree trains have fewer trunk passengers as their headway from the Ashmont train in front decreases. As Figure 9 shows, even headway allocation between branches, $H_{AB} = H_{BA} = 4.5$ min, is not the optimal allocation with respect to the overall transmission risk (assuming uniform infection rate in the service area). The overall infection probability reaches its lowest value at $H_{AB} = 5$ min and $H_{BA} = 4$ min. Changing the headway allocation can help mitigate transmission risk, lowering the overall infection probability by 3%.

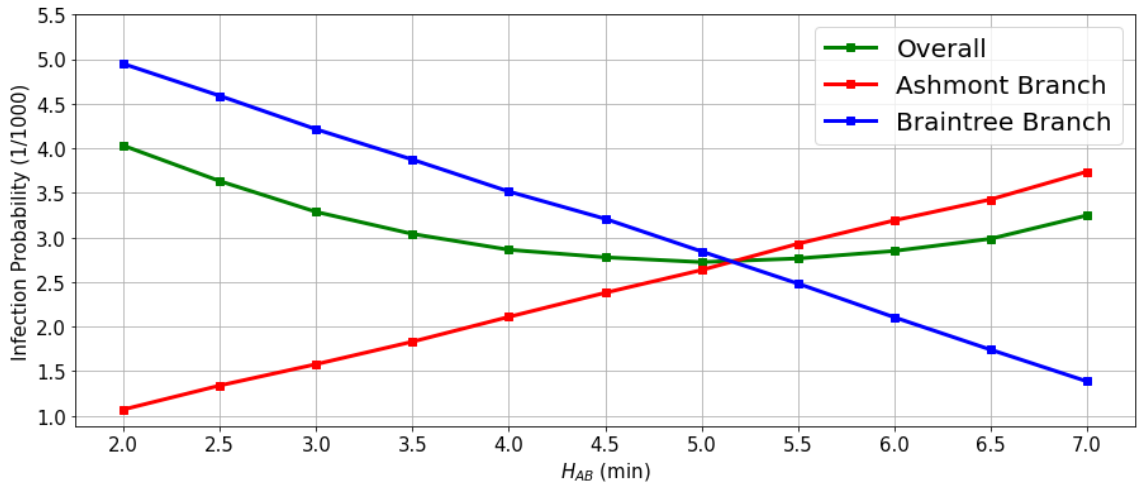

Figure 9 Infection probability under various headways between Ashmont and Braintree trains ($H_{AB}$)

Load on trains is impacted not only by OD flows and headways, but also by how passengers are distributed in train cars. When trains arrive at platforms, passengers tend to choose the location on the platform and consequently the boarding car, often based on the locations of the platform entrances and access locations (Jenelius and Cebecauer, 2020; Krstanoski, 2014). Uneven passenger distribution on platforms can cause uneven distribution in train cars. Passengers in crowded train cars experience higher transmission risk than passengers in other less crowded train cars. We analyze the impact on risk of the distribution of passengers among the various cars of a train using the Red Line setting. It is assumed that all stations in the network have similar patterns of passenger distribution. Each Red Line train has six cars. Based on results reported in Krstanoski (2014) for the passenger distribution at Bloor Station, Toronto, we consider two scenarios of car load distribution as shown in Figure 10. Under scenario 1, the 1st, 2nd 5th and 6th cars carry 12.5% of the passengers each and the 3rd and 4th cars carry 25% each. Under scenario 2, the 1st and 3rd cars carry 19% of the passengers, the 2nd car 24%, the 4th and 5th cars 16%, and the 6th car 6% (under even distribution each car carries 16.7% of the passengers).

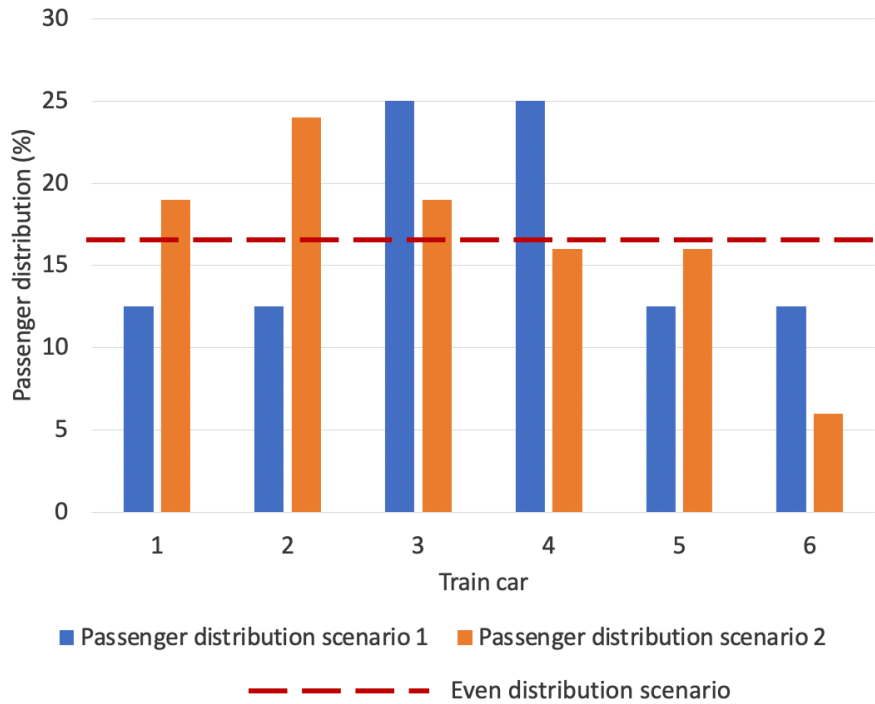

Figure 10 Passenger distribution among cars, scenarios 1 and 2, and even distribution

The results are shown in Figure 11. It is assumed that headways range from 4.0 to 11.0 mins and there are two levels of infection rate (0.1% and 1.0%). Figures 11a and 11d show that under passenger distribution scenarios 1 and 2, respectively, when the headways increase, overall system-wide infection probability increases significantly. However, the risk is similar under the two car load distributions. Figures 11b and 11e show the infection probability for the passengers who are on the most crowded car on each train. The infection probability for passengers on the most crowded car is more than 33% and 28% higher (scenarios 1 and 2 respectively) than the overall system-wide infection probability. Figures 11c and 11f show the infection probability for the passengers who are on the least crowded car. Not surprisingly, the least crowded train cars have a lower infection probability.

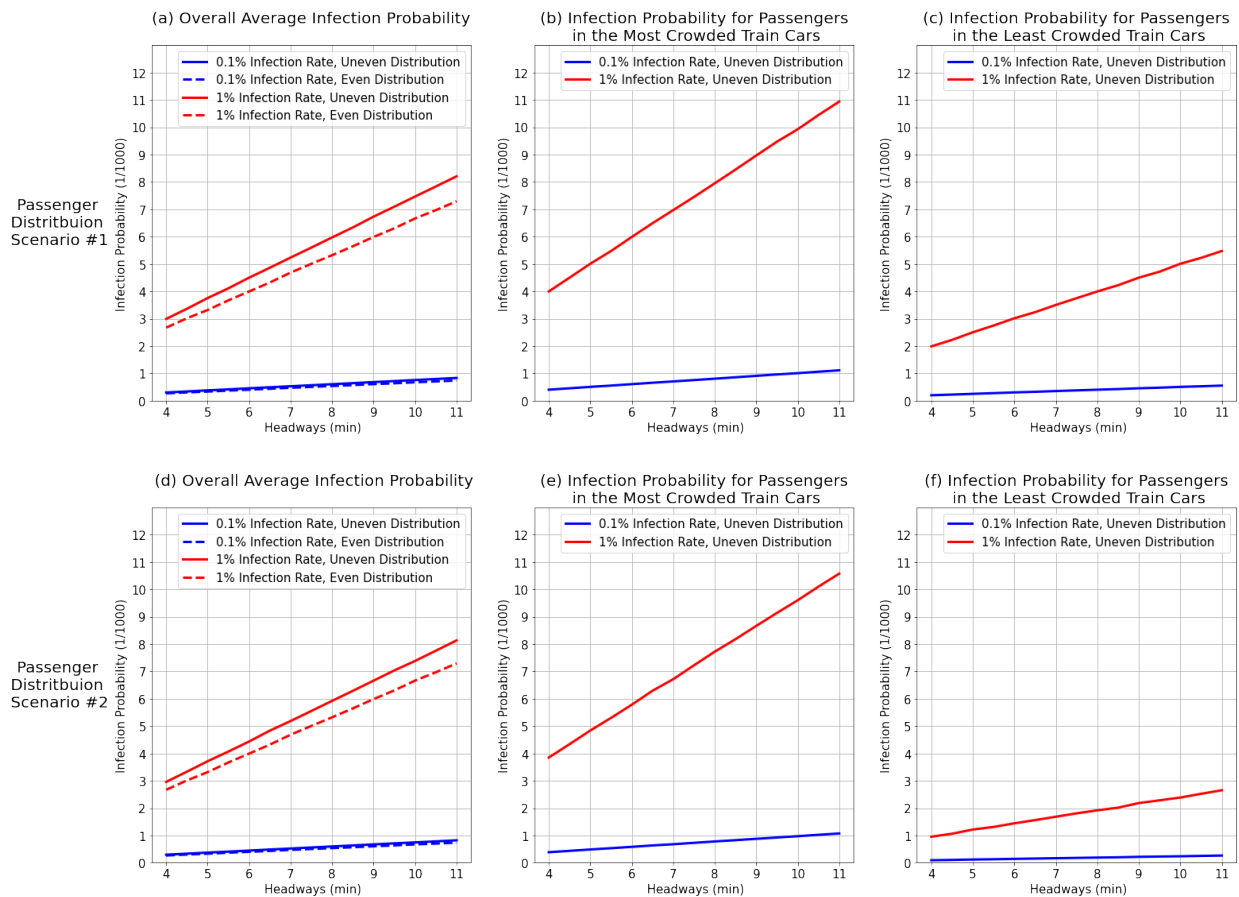

Figure 11 Infection probability for different passenger distribution scenarios under various dispatching headways and infection rates

It is important to point out that when the passenger distribution among train cars is non-uniform, as in the above cases, the overall system-wide infection probability can be up to 12.5% higher (depending on the conditions) than when trains have uniform passenger distribution among cars. At low infection rates (e.g. 0.1), infection risk is not sensitive to the load distribution.

Considering these findings, public transport agencies should actively manage the passenger distribution on the platforms during a pandemic to maintain relatively even passenger distribution among cars. The results also point out that other policies, where for example, an agency designates a car in the train only for elder and vulnerable users, may be worth examining.

## 4.3 Impact of Spatial Distribution of Infection Rates

The previous analyses assumed a uniform infection rate in the population across all stations. However, based on the COVID-19 infection rate data, the infection rates may vary by neighborhood or community, due to sociodemographic characteristics and other reasons (Dooling, 2020). Cambridge's infection rate on January 31, 2021 was 0.8% (City of Cambridge, 2021), Boston's 0.8% (City of Boston, 2021), and Quincy's 0.5% (City of Quincy, 2021). Ashmont's infection rates were not available. The positivity rate in the COVID-19 PCR (Polymerase Chain Reaction) tests can indicate the severity of infection rates. For example, Quincy's positivity rate was 6% and Boston's 9.7%, a relationship similar to the two cities' infection rates. The positivity rate for Ashmont was approximately 3 times of Quincy (Boston Public Health Commission, 2021; City of Quincy, 2021). Therefore, for the analysis it is assumed that the infection rate in Ashmont was 1.5%. The infection rate assumptions are shown in Figure 12.

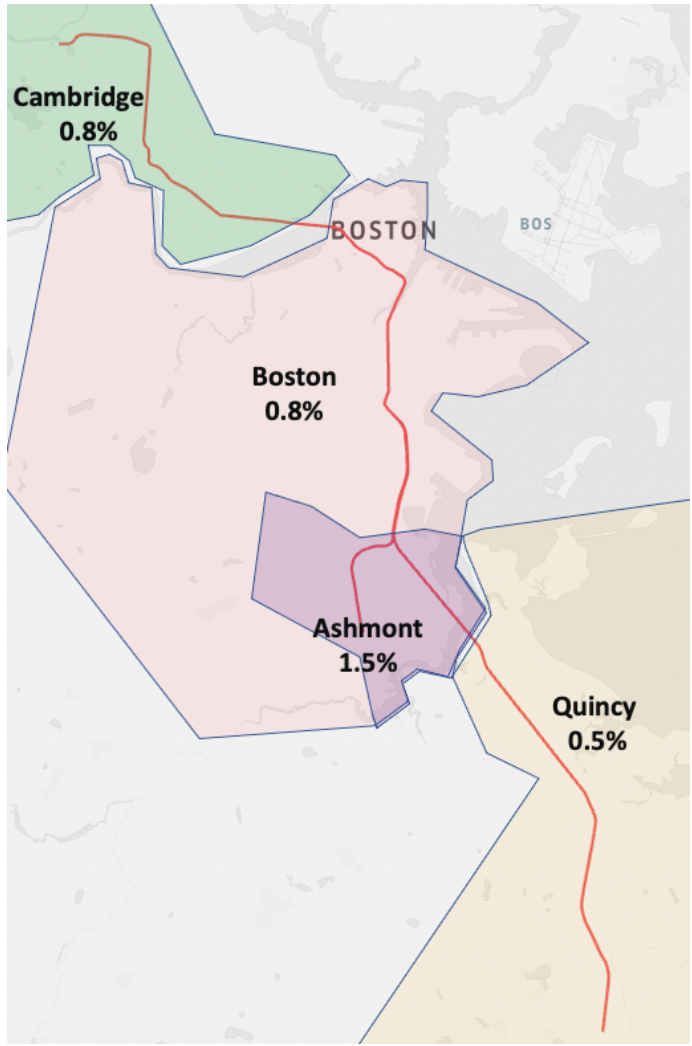

Figure 12 Infection rates in different municipalities and neighborhoods served by the Red Line

Two scenarios of spatial distribution of infection rates are studied. Scenario 1 assumes trunk passengers have a 0.8% infection rate, Braintree branch passengers 0.5%, and Ashmont branch passengers 1.5% as discussed above. In scenario 2, a uniform 0.92% infection rate is assumed. The value reflects the average infection rates used in scenario 1 weighted by the passenger demand in each section. The total expected number of carriers is the same as in scenario 1.

Under scenario 1, the spatial pattern of transmission risk for different OD pairs is shown in Figure 13. Figures 13a and c illustrate infection probabilities by OD pair while Figures 13b and d show the expected number of infections. Ashmont trains have a higher number of infections, even though they have lower demand and smaller train loads than Braintree trains. They also have a higher infection probability than Braintree trains.

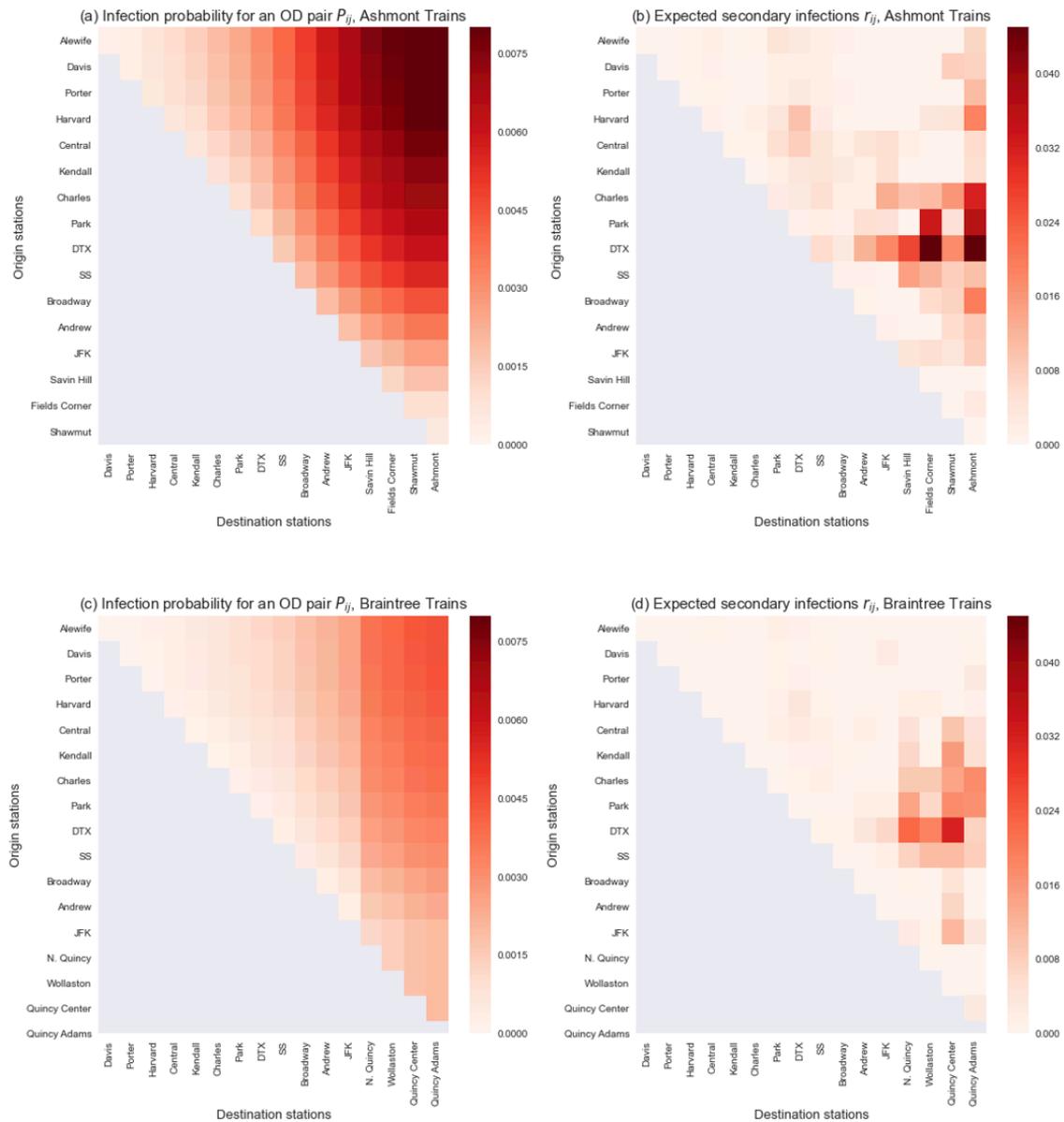

Figure 13 Infection probability and expected secondary infections (infection rate scenario 1)

Figure 14 shows results assuming the equivalent uniform infection rate of 0.92%. In this case, the determining factors for the risk distribution are the demand characteristics and the exposure time. OD pairs on the Braintree branch now have higher infection probabilities and more expected secondary infections, which is expected considering the higher train load and longer travel times (exposure). The system-wide probability of infection (overall transmission risk) under the more realistic spatial distribution of infection rates is 1.5% higher than under the uniform distribution.

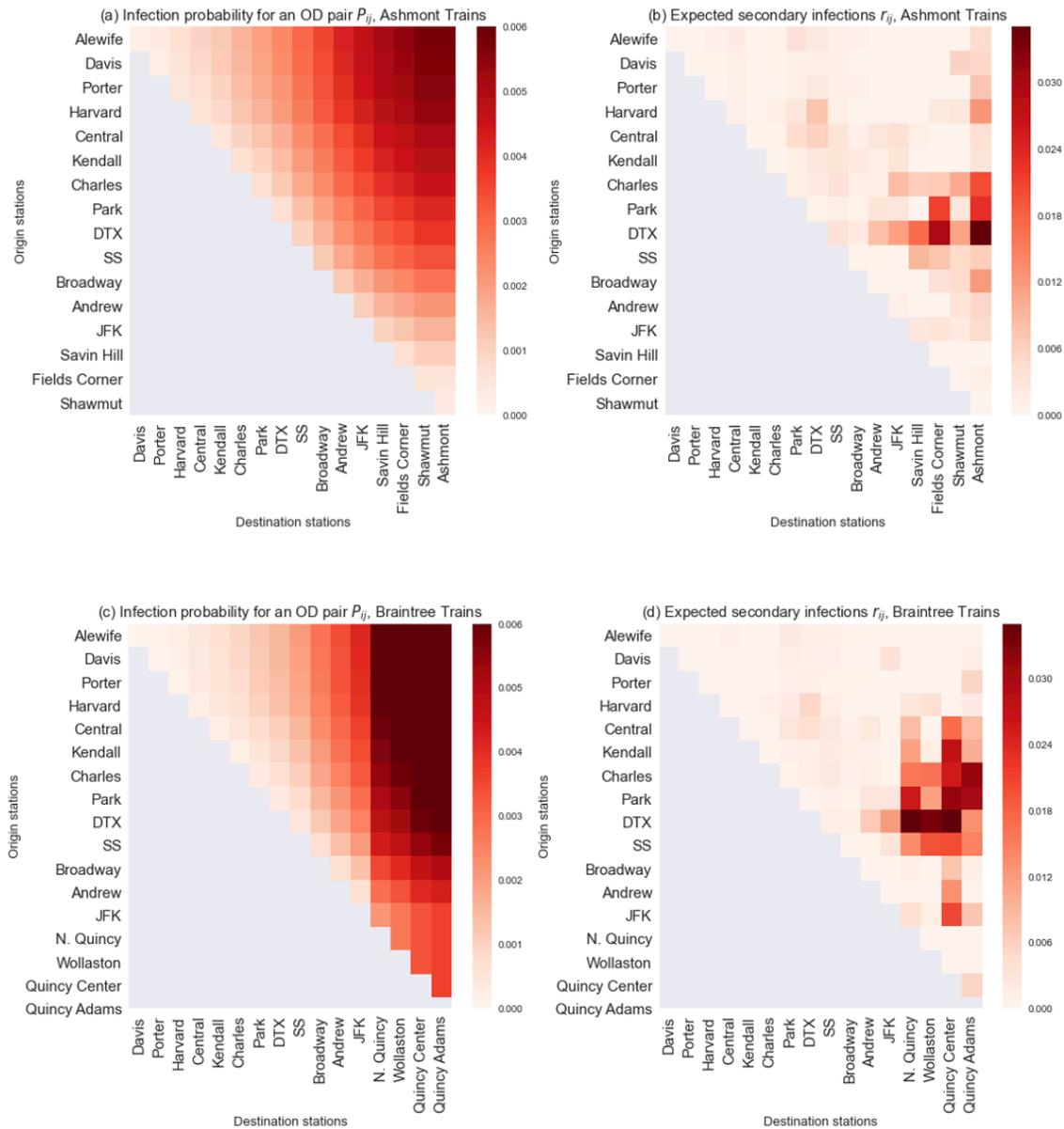

Figure 14 Infection probability and expected secondary infections (uniform infection rate, scenario 2)

To explore how spatially different infection rates impact the optimal headway allocation between the two branches, we analyze the risk assuming that the $H_{AB}$ headways range from 2 to 7 mins. As before, $H_{BA} = 9.0 - H_{AB}$. Figure 15a summarizes the results for the infection rate scenario 1 (actual infection rates). For comparison purposes, the results under infection rate scenario 2 (uniform infection rates) are also shown in Figure 15b. Based on the results shown in Figure 15a, the overall infection probability is lowest for $H_{AB} = 3$ min. Considering that the Ashmont branch has lower demand, the optimal allocation reduces the load on Ashmont trains even more, in order to mitigate the impact of higher infection rates in that branch. On the contrary, under scenario 2 with a uniform infection rate, the optimal allocation is $H_{AB} = 5.0$ min which reduces the load on the Braintree trains. The results further highlight the importance of considering the spatial patterns of demand and infection rates when evaluating mitigation strategies.

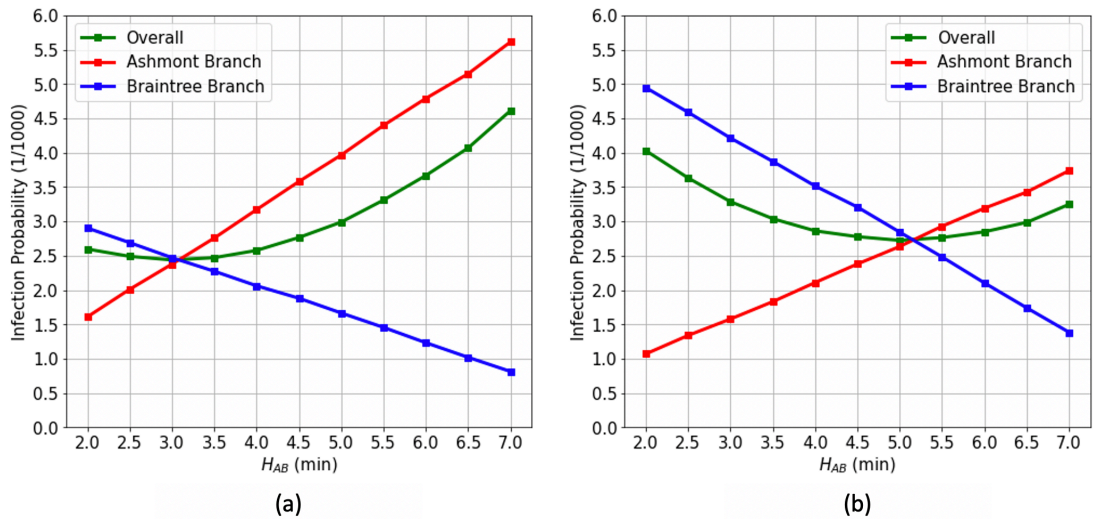

Figure 15 Infection probability as a function of headway allocation (a) non-uniform spatial distribution of infection rate; (b) equivalent uniform spatial distribution of infection rate

## 5. Conclusion

In this paper, we propose a virus transmission modeling framework based on the Wells-Riley model using as inputs transit operating characteristics, OD demand, and virus attributes. The model is sensitive to various factors that an operator can control, external factors related to the characteristics of the virus, as well as factors that may be subject to broader policy decisions (e.g. mask-wearing). It can be used to assess transmission risk as a function of OD flows, actual operations, and factors such as mask-wearing, ventilation, and magnitude and spatial characteristics of infection rates. Using data from the MBTA's Red Line, the study explores the effectiveness of different mitigation methods. Peak period risk increase is more significant with the increase in headways, when compared with the off-peak period. Also, mask-wearing is useful for risk mitigation. Increasing service frequency mitigates transmission risk. Adjusting the headway allocation to balance train loads in lines with branches can reduce the transmission risk. Furthermore, unbalanced passenger distribution on different cars on a train increases the overall system-wide infection risk. Spatial patterns, especially with respect to infection rates, are important and should be taken into account when designing strategies to mitigate risk. Possible directions for future work include extending the model to capture transmission risk at the platforms, so the overall risk associated with a trip is assessed.

**ACKNOWLEDGEMENTS**

The authors would like to thank the Massachusetts Bay Transportation Authority for their support for this research.

**REFERENCES**

Andrews, J.R., Morrow, C., Wood, R. (2013) Modeling the role of public transportation in sustaining tuberculosis transmission in South Africa. *American journal of epidemiology* 177, 556-561.

Barnett, A. (2020) Covid-19 Risk Among Airline Passengers: Should the Middle Seat Stay Empty? , MedRxiv.

Boston Public Health Commission (2021) COVID-19.

Buonanno, G., Morawska, L., Stabile, L. (2020a) Quantitative assessment of the risk of airborne transmission of SARS-CoV-2 infection: prospective and retrospective applications. *Environment International* 145, 106112.

Buonanno, G., Stabile, L., Morawska, L. (2020b) Estimation of airborne viral emission: Quanta emission rate of SARS-CoV-2 for infection risk assessment. *Environment International* 141, 105794.

Centers for Disease Control and Prevention (2021) Delta Variant: What We Know About the Science.

Chen, S.C., Liao, C.M., Li, S.S., You, S.H. (2011) A probabilistic transmission model to assess infection risk from Mycobacterium tuberculosis in commercial passenger trains. *Risk Analysis: An International Journal* 31, 930-939.

Chen, Y.-C., Lu, P.-E., Chang, C.-S., Liu, T.-H. (2020) A time-dependent SIR model for COVID-19 with undetectable infected persons. *IEEE Transactions on Network Science and Engineering* 7, 3279-3294.

City of Boston (2021) COVID-19 Case Tracker.

City of Cambridge (2021) Cambridge COVID-19 Data Center.

City of Quincy (2021) City of Quincy COVID-19 Information.

Coronaviridae Study Group of the International Committee on Taxonomy of Viruses (2020) The species Severe acute respiratory syndrome-related coronavirus: classifying 2019-nCoV and naming it SARS-CoV-2. *Nature microbiology* 5, 536.

Dai, H., Zhao, B. (2020) Association of the infection probability of COVID-19 with ventilation rates in confined spaces. *Proceedings of Building Simulation*, pp. 1321-1327.

American Public Transportation Association (2020) *Public Transportation Ridership Report: Second Quarter 2020.*

Dooling, S. (2020) Why Some Boston Neighborhoods Have Been Hit Harder By The Pandemic Than Others. WBUR.

Environmental Protection Agency (2015) Chapter 6., *Exposure Factors Handbook 2011 Edition (Final Report)* ed Environmental Protection Agency.

Fennelly, K.P., Nardell, E.A. (1998) The relative efficacy of respirators and room ventilation in preventing occupational tuberculosis. *Infection Control & Hospital Epidemiology* 19, 754-759.

Furuya, H. (2007) Risk of transmission of airborne infection during train commute based on mathematical model. *Environmental health and preventive medicine* 12, 78-83.

Goscé, L., Johansson, A. (2018) Analysing the link between public transport use and airborne transmission: mobility and contagion in the London underground. *Environmental Health* 17, 1-11.

Harris, J.E. (2020) The subways seeded the massive coronavirus epidemic in New York City. *NBER working paper.*

Herrero, L. (2021) How contagious is Delta? How long are you infectious? Is it more deadly? A quick guide to the latest science.

Congressional Reserach Service (2020) *COVID-19: State and Local Shut-Down Orders and Exemptions for Critical Infrastructure.*

Jenelius, E., Cebecauer, M. (2020) Impacts of COVID-19 on public transport ridership in Sweden: Analysis of ticket validations, sales and passenger counts. *Transportation Research Interdisciplinary Perspectives* 8, 100242.

Ko, G., Thompson, K.M., Nardell, E.A. (2004) Estimation of tuberculosis risk on a commercial airliner. *Risk Analysis: An International Journal* 24, 379-388.

Koutsopoulos, H.N., Wang, Z. (2007) Simulation of urban rail operations: Application framework. *Transportation research record* 2006, 84-91.

Krstanoski, N. (2014) MODELLING PASSENGER DISTRIBUTION ON METRO STATION PLATFORM. *International Journal for Traffic & Transport Engineering* 4.

Liu, L., Miller, H.J., Scheff, J. (2020) The impacts of COVID-19 pandemic on public transit demand in the United States. *Plos one* 15, e0242476.

Luo, K., Zheng, H., Xiao, S., Yang, H., Jing, X., Wang, H., Xie, Z., Luo, P., Li, W., Li, Q., Tan, H., Xu, Z., Hu, S. (2020) An epidemiological investigation of 2019 novel coronavirus disease through aerosol-borne transmission by public transport. *Practical Preventive Medicine* 27.

Massachusetts Bay Transportation Authority (2013) Red and Orange Line New Vehicle Procurement.

Massachusetts Bay Transportation Authority (2019) MBTA Performance dashboard: Ridership. 2019.

Miller, S.L., Nazaroff, W.W., Jimenez, J.L., Boerstra, A., Buonanno, G., Dancer, S.J., Kurnitski, J., Marr, L.C., Morawska, L., Noakes, C. (2021) Transmission of SARS-CoV-2 by inhalation of respiratory aerosol in the Skagit Valley Chorale superspreading event. *Indoor air* 31, 314-323.

Mo, B., Feng, K., Shen, Y., Tam, C., Li, D., Yin, Y., Zhao, J. (2021) Modeling epidemic spreading through public transit using time-varying encounter network. *Transportation Research Part C: Emerging Technologies* 122, 102893.

Orro, A., Novales, M., Monteagudo, Á., Pérez-López, J.-B., Bugarín, M.R. (2020) Impact on City Bus Transit Services of the COVID–19 Lockdown and Return to the New Normal: The Case of A Coruña (Spain). 12, 7206.

Pan, J., Harb, C., Leng, W., Marr, L.C. (2020) Inward and outward effectiveness of cloth masks, a surgical mask, and a face shield. medRxiv, p. 2020.2011.2018.20233353.

Park, J. (2020) Changes in subway ridership in response to COVID-19 in Seoul, South Korea: Implications for social distancing. *Cureus* 12.

Prem, K., Liu, Y., Russell, T.W., Kucharski, A.J., Eggo, R.M., Davies, N., Flasche, S., Clifford, S., Pearson, C.A., Munday, J.D. (2020) The effect of control strategies to reduce social mixing on outcomes of the COVID-19 epidemic in Wuhan, China: a modelling study. *The Lancet Public Health* 5, e261-e270.

Riley, E., Murphy, G., Riley, R. (1978) Airborne spread of measles in a suburban elementary school. *American journal of epidemiology* 107, 421-432.

Rudnick, S., Milton, D. (2003) Risk of indoor airborne infection transmission estimated from carbon dioxide concentration. *Indoor air* 13, 237-245.

Shen, Y., Li, C., Dong, H., Wang, Z., Martinez, L., Sun, Z., Handel, A., Chen, Z., Chen, E., Ebell, M.H. (2020) Community outbreak investigation of SARS-CoV-2 transmission among bus riders in eastern China. *JAMA internal medicine* 180, 1665-1671.

(2012) *HVAC filtration and the Wells-Riley approach to assessing risks of infectious airborne diseases*.

Stutt, R.O., Retkute, R., Bradley, M., Gilligan, C.A., Colvin, J. (2020) A modelling framework to assess the likely effectiveness of facemasks in combination with 'lock-down'in managing the COVID-19 pandemic. *Proceedings of the Royal Society A* 476, 20200376.

Tirupathi, R., Bharathidasan, K., Palabindala, V., Salim, S.A., Al-Tawfiq, J.A. (2020) Comprehensive review of mask utility and challenges during the COVID-19 pandemic. *Infez Med* 28, 57-63.

Zhou, J., Koutsopoulos, H.N. (2020) Virus Transmission Risk in Urban Rail Systems: A Microscopic Simulation-based Analysis of Spatio-temporal Characteristics. *arXiv:2008.08448*. arXiv.

Zhou, J., Koutsopoulos, H.N. (2021) Virus Transmission Risk in Urban Rail Systems: Microscopic Simulation-Based Analysis of Spatio-Temporal Characteristics. *Transportation Research Record*, 03611981211010181.

Zhou, J., Koutsopoulos, H.N., Saidi, S. (2020) Evaluation of Subway Bottleneck Mitigation Strategies using Microscopic, Agent-Based Simulation. *Transportation Research Record* 2674, 649-661.

Zhu, R., Anselin, L., Batty, M., Kwan, M.-P., Chen, M., Luo, W., Cheng, T., Lim, C.K., Santi, P., Cheng, C., Gu, Q., Wong, M.S., Zhang, K., Lü, G., Ratti, C. (2021) The effects of different travel modes and travel destinations on COVID-19 transmission in global cities. *Sci Bull (Beijing)*.